\documentclass[12pt]{article}
\usepackage{amsmath,amssymb}
\usepackage{bm}
\usepackage{color}
\usepackage{graphicx}
\usepackage{cite}

\begin{document}

\title{
\begin{flushright}
\ \\*[-80pt]
\begin{minipage}{0.2\linewidth}
\normalsize
EPHOU-19-013\\
HUPD1913 \\*[50pt]
\end{minipage}
\end{flushright}
{\Large \bf
$CP$ violation in modular invariant flavor models
\\*[20pt]}}

\author{
Tatsuo Kobayashi $^{1}$
,~Yusuke Shimizu $^{2}$
,~Kenta Takagi $^{2}$
,\\Morimitsu Tanimoto $^{3}$
,~Takuya H. Tatsuishi $^{1}$
, and~Hikaru Uchida $^{1}$
\\*[20pt]
\centerline{
\begin{minipage}{\linewidth}
\begin{center}
$^1${\it \normalsize
Department of Physics, Hokkaido University, Sapporo 060-0810, Japan} \\*[5pt]
$^2${\it \normalsize
Graduate School of Science, Hiroshima University, Higashi-Hiroshima 739-8526, Japan}\\*[5pt]
$^3${\it \normalsize
Department of Physics, Niigata University, Niigata 950-2181, Japan}
\end{center}
\end{minipage}}
\\*[50pt]}

\date{
\centerline{\small \bf Abstract}
\begin{minipage}{0.9\linewidth}
\medskip
\medskip
\small
We study the spontaneous $CP$ violation through the stabilization of the modulus $\tau$ in 
modular invariant flavor models.
The $CP$-invariant potential has the minimum only at ${\rm Re}[\tau] = 0$ or 1/2  (mod 1).
From this result, we study $CP$ violation in modular invariant flavor models.
The physical $CP$ phase is vanishing.
The important point for the $CP$ conservation is the $T$ transformation in the 
modular symmetry.
One needs the violation of $T$ symmetry to realize the $CP$ violation.
\end{minipage}
}

\begin{titlepage}
\maketitle
\thispagestyle{empty}
\end{titlepage}
\newpage
\section{Introduction}

Particle physics still has several mysteries.
Mysteries related to the flavor origin are one of important issues to study, i.e.
the family number, the hierarchy of quark and lepton masses, their mixing angles, and the origin of the $CP$ violation.

Various studies have been carried out to solve these flavor mysteries.
One of the interesting approaches is non-Abelian discrete flavor symmetries \cite{
	Altarelli:2010gt,Ishimori:2010au,Ishimori:2012zz,Hernandez:2012ra,
	King:2013eh,King:2014nza,Tanimoto:2015nfa,King:2017guk,Petcov:2017ggy}.
In this approach, a non-Abelian discrete flavor symmetry is imposed and its nontrivial representations are assigned to three families of quarks and leptons.
These flavor symmetries are broken by vacuum expectation values of scalar fields, the so-called flavon fields such that one realizes quark and lepton masses and their mixing angles and $CP$ phases.
A great deal of models have been constructed by use of various discrete groups such as $S_N$, $A_N$, $\Delta(3N^2)$, $\Delta(6N^2)$, etc. for quarks and leptons.

An underlying theory may have an origin for these flavor symmetries.
Within the framework of extra dimensional field theory and superstring theory, 
geometrical symmetries of compact space can provide us with the origin of these flavor symmetries.\footnote{
  See Refs.~\cite{Kobayashi:2006wq,Kobayashi:2004ya,Ko:2007dz,Beye:2014nxa,Abe:2009vi} for non-Abelian discrete flavor symmetries in superstring theory.}
Torus and orbifold compactifications are simple compactifications, and these compactifications have the geometrical symmetry, i.e.
the so-called modular symmetry $SL(2,\mathbb{Z})$, which corresponds to the change of the torus basis.
The ratio of basis vectors is denoted by the modulus $\tau$, that is, the modulus describes the shape of the torus and the orbifold.
The modular group transforms the modulus $\tau$ nontrivially.
Yukawa couplings and other couplings depend on the modulus.
Thus, the modular group transforms these couplings nontrivially.
Note that zero modes corresponding to quarks and leptons transform each other under the modular group within the framework of superstring theory \cite{
	Lauer:1989ax,Lerche:1989cs,Ferrara:1989qb,
	Kobayashi:2017dyu,Kobayashi:2018rad,Baur:2019kwi,Kariyazono:2019ehj}.
The modular group is in this sense a flavor symmetry.
It is further interesting that quotient groups, $\Gamma_N = \overline \Gamma/\overline \Gamma (N)$ are isomorphic to 
$\Gamma_2 \simeq S_3$, $\Gamma_3 \simeq A_4$, $\Gamma_4 \simeq S_4$ and $\Gamma_5 \simeq A_5$ 
\cite{deAdelhartToorop:2011re}.

Recently, inspired by these aspects, a new approach of flavor model building was proposed in Ref.~\cite{Feruglio:2017spp}.
The $A_4$ modular symmetry is assumed in Ref.~\cite{Feruglio:2017spp}.
The three families of leptons are assigned to nontrivial representations.
The couplings are nontrivial representations of the $A_4$ modular symmetry and depend on the modulus.
The couplings are modular forms and transform under $A_4$ in this scenario.
We can construct models which lead to a realistic result \cite{Criado:2018thu,Kobayashi:2018scp} by fixing appropriate values of the modulus $\tau$ and other parameters.
The modular forms of weight 2 are fundamental and modular forms of higher weights can be obtained by their products.
In the past year, such fundamental modular forms of weight 2 have been constructed for 
$S_3$ \cite{Kobayashi:2018vbk},
$S_4$ \cite{Penedo:2018nmg},
$A_5$ \cite{Novichkov:2018nkm},
$\Delta(96)$, and $\Delta(384)$ \cite{Kobayashi:2018bff}.
Modular forms of odd weights are possible for double covering groups.
For example, the modular forms of the weight 1 and higher weights are also given for the $T'$ doublet \cite{Liu:2019khw}.
The new approach of model building to the flavors, i.e. the flavor modular symmetric models, has been studied by use of these modular forms\cite{
	Criado:2018thu,Kobayashi:2018scp,Novichkov:2018ovf,
	deAnda:2018ecu,Okada:2018yrn,Kobayashi:2018wkl,Novichkov:2018yse,Ding:2019xna,Nomura:2019jxj,Novichkov:2019sqv,
	Okada:2019uoy,deMedeirosVarzielas:2019cyj,Nomura:2019yft,Kobayashi:2019rzp,Okada:2019xqk,Kobayashi:2019mna,
	Ding:2019zxk,Okada:2019mjf,King:2019vhv,Nomura:2019lnr,Okada:2019lzv,Criado:2019tzk,Gui-JunDing:2019wap,
Zhang:2019ngf,Wang:2019ovr}.

One of the important features in these flavor modular symmetric models is that the flavor symmetry is broken when the value of the modulus $\tau$ is fixed.
Thus, one does not need flavon fields to break flavor symmetries.
Fixing the value of $\tau$ is one of most important issues, which is called the modulus stabilization.
The modulus stabilization was studied within the framework of supergravity theory.
One can find the modular invariance of supergravity theory in the literature \cite{Ferrara:1989bc}\footnote{
	See, e.g. \cite{Derendinger:1991hq,Ibanez:1992hc,Kobayashi:2016ovu}, for their applications.}.
The modulus stabilization was studied by assuming the $SL(2,\mathbb{Z})$ modular invariance for the nonperturbative superpotential in supergravity theory \cite{Ferrara:1990ei,Cvetic:1991qm} \footnote{
	See also \cite{Kobayashi:2016mzg}.}.
Recently, such analysis was extended to one of the flavor modular symmetric models \cite{Kobayashi:2019xvz}.

Higher dimensional theories such as higher dimensional super Yang-Mills theory and superstring theory conserve $CP$.
The four-dimensional $CP$ symmetry can be embedded into $(4+d)$ dimensions as higher dimensional proper Lorentz symmetry with positive determinant \cite{
  Green:1987mn,Strominger:1985it,Dine:1992ya,Choi:1992xp,Lim:1990bp,Kobayashi:1994ks}.
That is, one can combine the four-dimensional $CP$ transformation and $d$-dimensional transformation with negative determinant so as to obtain $(4+d)$ dimensional proper Lorentz transformation.
For example in six-dimensional theory, we denote the two extra coordinates by a complex coordinate $z$.
The four-dimensional $CP$ symmetry with $z \rightarrow z^*$ or $z \rightarrow -z^*$ is a six-dimensional proper Lorentz symmetry.
(See for \cite{Novichkov:2019sqv} $CP$ symmetry from the viewpoint of flavor modular symmetries.)
Extensions to other dimensions are straightforward.
Thus, breaking of such a symmetry corresponds to the $CP$ violation.
The above transformation of the coordinate $z$ corresponds to the transformation of the modulus.

The modulus stabilization fixes the modulus value.
The $CP$ violation can occur through the modulus stabilization \cite{Acharya:1995ag,Dent:2001cc,Khalil:2001dr,Giedt:2002ns}.
The purpose of this paper is to study the spontaneous $CP$ violation through the modulus stabilization in flavor modular symmetric models.

This paper is organized as follows.
In section 2, we give a brief review on the modular symmetry and the $CP$ symmetry, which is embedded in higher dimensions.
We study the modulus stabilization in the $CP$-invariant scalar potential in section 3.
It is shown that the minimum of the $CP$-invariant potential corresponds to ${\rm Re}[\tau] =0$ or $1/2$  (mod 1).
We study the $CP$  at ${\rm Re}[\tau] =1/2$ in an $A_4$ flavor model, and study generic aspect in section 4.
Section 5 is conclusion.
Appendices A and B show modular forms of the levels 3 and 4, respectively.

\section{Modular symmetry and $CP$}
\subsection{Modular symmetry}

We briefly review the modular symmetry.
The two-dimensional torus is constructed by $\mathbb{R}^2/\Lambda$, where $\Lambda$ is a two-dimensional lattice.
The lattice itself is spanned by two basis vectors, $\alpha_1$ and $\alpha_2$.
We denote them as $\alpha_1=2 \pi R$ and $\alpha_2 = 2 \pi R\tau$, where $R$ is real and $\tau$ is complex.
One can use another basis to span the same lattice $\Lambda$, that is, the same lattice $\Lambda$ is spanned by the following two lattice vectors, 
\begin{equation}
  \label{eq:SL2Z}
  \left(
  \begin{array}{c}
  \alpha'_2 \\ \alpha'_1
  \end{array}
  \right) =\left(
  \begin{array}{cc}
  a & b \\
  c & d 
  \end{array}
  \right) \left(
  \begin{array}{c}
  \alpha_2 \\ \alpha_1
  \end{array}
  \right) \ ,
\end{equation}
where $a,b,c,d$ are integers with satisfying $ad-bc = 1$.
That is the $SL(2,\mathbb{Z})$ transformation.

The modulus $\tau = \alpha_2/\alpha_1$ transforms under the above change of bases,
\begin{equation}\label{eq:tau-SL2Z}
  \tau \longrightarrow \tau'=\gamma \tau = \frac{a\tau + b}{c \tau + d}\ .
\end{equation}
This is the modular symmetry.
Note that the modulus transforms identically, $\tau \rightarrow \tau$, under the $\mathbb{Z}_2$ transformation, i.e. $a=d=-1$ and $b=c=0$.
The symmetry is therefore $PSL(2,\mathbb{Z})=SL(2,\mathbb{Z})/\mathbb{Z}_2$, which is denoted by $\bar \Gamma$.
We restrict ourselves to the upper half plane for $\tau$: ${\rm Im}[\tau] >0$.

The modular group has two generators, $S$ and $T$, which transform 
\begin{equation}
	S: \tau \longrightarrow -\frac{1}{\tau},\qquad 
	T: \tau \longrightarrow \tau +1.
\end{equation}
They satisfy the following algebraic relations,
\begin{equation}
	\label{eq:S-ST}
	S^2=(ST)^3=\mathbb{I}.
\end{equation}
The congruence subgroups of level $N$ are defined as 
\begin{equation}
	\bar \Gamma(N) = \left\{ \left( 
	\begin{array}{cc} a & b \\ c & d 
	\end{array}
	\right) \in PSL(2,\mathbb{Z}), \qquad \left( 
	\begin{array}{cc} a & b \\ c & d 
	\end{array}
	\right) = \left( 
	\begin{array}{cc} 1 & 0 \\ 0 & 1 
	\end{array}
	\right) \quad ({\rm mod}~N) \right\}.
\end{equation}
Furthermore, the quotient groups $\Gamma_N$ are given as $\Gamma_N\equiv \bar \Gamma/\bar \Gamma(N)$.
These are finite groups for $N=2,3,4,5$, and isomorphic to $A_N$ or $S_N$:
$\Gamma_2 \simeq S_3$, $\Gamma_3 \simeq A_4$, $\Gamma_4\simeq S_4$, $\Gamma_5 \simeq A_5$,
where the algebraic relation $T^N = \mathbb{I}$ is satisfied in addition to Eq.(\ref{eq:S-ST}).

The modular forms of weight $k$ are the holomorphic functions of $\tau$ and transform as
\begin{equation}
	f_i(\tau) \longrightarrow (c\tau +d)^k \rho(\gamma)_{ij}f_j( \tau),
\end{equation}
under the modular symmetry, where $\rho(\gamma)_{ij}$ is a unitary matrix under $\Gamma_N$.
The chiral matter fields also transform nontrivially 
\begin{equation}
	(\phi^{(I)})_i(x) \longrightarrow (c\tau +d)^{-k_I} \rho(\gamma)_{ij}(\phi^{(I)})_j(x).
\end{equation}

For example, there are three modular forms of the level 3 and the weight 2 for $\bar \Gamma(3)$.
They can be written by \cite{Feruglio:2017spp},
\begin{eqnarray} 
  \label{eq:Y-A4}
  Y_1^{3,2}(\tau) &=& \frac{i}{2\pi}\left( \frac{\eta'(\tau/3)}{\eta(\tau/3)} +\frac{\eta'((\tau +1)/3)}{\eta((\tau+1)/3)} 
  +\frac{\eta'((\tau +2)/3)}{\eta((\tau+2)/3)} - \frac{27\eta'(3\tau)}{\eta(3\tau)} \right), \nonumber \\
  Y_2^{3,2}(\tau) &=& \frac{-i}{\pi}\left( \frac{\eta'(\tau/3)}{\eta(\tau/3)} +\omega^2\frac{\eta'((\tau +1)/3)}{\eta((\tau+1)/3)} 
  +\omega \frac{\eta'((\tau +2)/3)}{\eta((\tau+2)/3)} \right) , \label{tripletY} \\ 
  Y_3^{3,2}(\tau) &=& \frac{-i}{\pi}\left( \frac{\eta'(\tau/3)}{\eta(\tau/3)} +\omega\frac{\eta'((\tau +1)/3)}{\eta((\tau+1)/3)} 
  +\omega^2 \frac{\eta'((\tau +2)/3)}{\eta((\tau+2)/3)} \right)\,,
  \nonumber
  \label{YA4}
\end{eqnarray}
where $q = e^{2 \pi i \tau}$ and $\omega=e^{i\frac{2}{3}\pi}$, and 
$\eta(\tau)$ denotes the Dedekind eta function: 
\begin{equation}
  \eta(\tau) = q^{1/24} \prod_{n =1}^\infty (1-q^n)~.
\end{equation}
They are triplet under $A_4 \simeq \Gamma_3$, where $S$ and $T$ are represented by
\begin{align}
  \begin{aligned}
  \rho(S)=\frac{1}{3}
  \begin{pmatrix}
  -1 & 2 & 2 \\
  2 &-1 & 2 \\
  2 & 2 &-1
  \end{pmatrix},
  \end{aligned}
  \qquad 
  \begin{aligned}
  \rho(T)=
  \begin{pmatrix}
  1 & 0& 0 \\
  0 &\omega& 0 \\
  0 & 0 & \omega^2
  \end{pmatrix}.
  \end{aligned}
\end{align}
These modular forms are expanded by $q$ as 
\begin{align}
  Y^{3,2}_{\bf 3}=\begin{pmatrix}Y_1^{3,2}(\tau)\\
  Y_2^{3,2}(\tau)\\
  Y_3^{3,2}(\tau)\end{pmatrix}=
  \begin{pmatrix}
  1+12q+36q^2+12q^3+\dots \\
  -6q^{1/3}(1+7q+8q^2+\dots) \\
  -18q^{2/3}(1+2q+5q^2+\dots)\end{pmatrix}.
  \label{q-expansion}
\end{align}
The modular forms of higher weights are obtained by their products.
For example, the $A_4$ trivial-singlet modular form of the weight 4 can be written by
\begin{equation}
  Y^{3,4}_{\bf 1}= (Y_1^{3,2})^2+2Y_2^{3,2}Y_3^{3,2}.
\end{equation}
Other modular forms of the level 3 and the weight 4 are shown in Appendix A.

There are two modular forms of the level 2 and the weight 2 for $\bar \Gamma(2)$.
These can be written by \cite{Kobayashi:2018vbk},
\begin{eqnarray} 
  \label{eq:Y-S3}
  Y_{1}^{2,2}(\tau) &=& \frac{i}{4\pi}\left( \frac{\eta'(\tau/2)}{\eta(\tau/2)} +\frac{\eta'((\tau +1)/2)}{\eta((\tau+1)/2)} 
  - \frac{8\eta'(2\tau)}{\eta(2\tau)} \right), \nonumber \\
  Y_2^{2,2}(\tau) &=& \frac{\sqrt{3}i}{4\pi}\left( \frac{\eta'(\tau/2)}{\eta(\tau/2)} -\frac{\eta'((\tau +1)/2)}{\eta((\tau+1)/2)} \right) . \label{doubletY} \nonumber
\end{eqnarray}
They are doublet under $S_3 \simeq \Gamma_2$, where $S$ and $T$ are represented by 
\begin{equation}
  \rho(S) = \frac{1}{2}\left(
  \begin{array}{cc}
  -1 & -\sqrt{3} \\
  -\sqrt{3} & 1
  \end{array}\right), \qquad\qquad 
  \rho(T) = \left(
  \begin{array}{cc}
  1 & 0 \\
  0 & -1
  \end{array}\right).
  \label{S3base}
\end{equation}
They can be expanded by $q$, 
\begin{align}
  Y^{2,2}_{\bf 2}=\begin{pmatrix}Y_1^{2,2}(\tau)\\Y_2^{2,2}(\tau)
  \end{pmatrix}=
  \begin{pmatrix}
  \frac{1}{8}+3q+3q^2+12q^3+3q^4+\dots \\
  \sqrt{3}q^{1/2}(1+4q+6q^2+8q^3+\dots ) \end{pmatrix}.
\end{align}
Furthermore, the $S_3$ trivial singlet of the weight 4 can be written by 
\begin{equation} 
  Y^{2,4}_{\bf 1} = (Y^{2,2}_1(\tau))^2+(Y^{2,2}_2(\tau))^2 ~ .
\end{equation}

The five modular forms of the level 4 and weight 2 for $\bar \Gamma(4)$ are found in Ref.~\cite{Penedo:2018nmg}.
These are shown in Appendix B.
They correspond to ${\bf 2}$ and ${\bf 3}'$ under $S_4$.
The $S_4$ trivial singlet of the weight 4 can be written by $q$-expansion \cite{Novichkov:2019sqv}, 
\begin{equation}
  Y^{4,4}_{\bf 1} = \frac{1}{64}+ \frac{15}{4}q+ \frac{135}{4}q^2+135 q^3 + \cdots,
\end{equation}
up to an overall factor.

\subsection{$CP$}

The four-dimensional $CP$ can be embedded to higher dimensional symmetry.
We focus on the six dimensions here.
A six-dimensional proper Lorentz symmetry is the combination of the four-dimensional $CP$ and a transformation with negative determinant in the extra two dimensions.
Such extra dimensional transformations correspond to $z \rightarrow z^*$ and  $z \rightarrow -z^*$.
Note that $z = x + \tau y$, where $x$ and $y$ are real coordinates.
The latter transformation $z \rightarrow -z^*$ maps the upper half plane ${\rm Im}[\tau]>0$ to the same half plane.
Hence, we consider the transformation $z \rightarrow -z^*$ as the $CP$ symmetry.
That means that the $CP$ transforms the modulus $\tau$ 
\begin{equation}\label{eq:CP-tau}
  \tau \rightarrow -\tau^* .
\end{equation}
The same transformation of $\tau$ was derived from the viewpoint of generalized $CP$ symmetry embedded in modular symmetry from the viewpoint of superstring theory \cite{Baur:2019kwi} and four-dimensional models \cite{Novichkov:2019sqv}.\footnote{
  See for generalized $CP$ \cite{Feruglio:2012cw,Holthausen:2012dk,Chen:2014tpa}.}
We study the modulus stabilization in the next section.
Once the value of $\tau$ is fixed at the generic point, the above symmetry (\ref{eq:CP-tau}) is broken.
Thus, the $CP$ can also be spontaneously violated through the modulus stabilization.
However,  ${\rm Re}[\tau]=0$ is a symmetric line.
Also ${\rm Re}[\tau]=1/2$  (mod 1) is another symmetric line up to modular transformation, 
because $\tau = 1/2+i {\rm Im}[\tau] \rightarrow -\tau^*= -(1/2)+i {\rm Im}[\tau]=\tau -1$.\footnote{
In addition, $CP$ is conserved along the curve $|\tau|=1$ up to the modular symmetry, because 
$\tau = e^{ia} \rightarrow -\tau^*= -e^{-ia}= -1/\tau$.}
Also ,the $CP$ conservation at ${\rm Re}[\tau]=1/2$ will  be studied explicitly in Yukawa matrices later, 
after study on the modulus stabilization.

\section{Modulus stabilization}

We study the modulus stabilization within the framework of supergravity theory following 
Refs.~\cite{Ferrara:1990ei,Cvetic:1991qm,Kobayashi:2019xvz}.
We focus only on the $CP$ violation, that is, the value of ${\rm Re}[\tau]$.
We use the unit $M_P=1$.

Supergravity theory Lagrangian can be written by $G$, 
\begin{equation}
	G = K + \ln |W|^2,
\end{equation}
where $K$ and $W$ denote the K\"{a}hler potential and the superpotential.
The scalar potential is written by 
\begin{equation}
  V= e^G(G^{-1}_{\tau \bar \tau} |G_\tau|^2 -3) = e^K(K^{-1}_{\tau \bar \tau} |D_\tau W|^2 - 3|W|^2),
\end{equation}
where 
\begin{equation}
	\label{eq:DW=0}
	D_\tau W = K_\tau W + W_\tau,
\end{equation}
with $G_\tau = \partial G/\partial \tau$, $K_\tau = \partial K/\partial \tau$ and $W_\tau = \partial W/ \partial \tau$.

The typical K\"ahler potential of the modulus field $\tau$ is obtained as 
\begin{equation}
	\label{eq:Kahler-tau}
	K = - \ln [i(\bar\tau - \tau)], 
\end{equation}
and it transforms as 
\begin{equation}
	-\ln[i(\bar\tau - \tau)] \longrightarrow -\ln[i(\bar\tau - \tau)] + \ln |c\tau + d|^2,
\end{equation}
under the modular transformation.
The superpotential should transform as
\begin{equation}
	W \longrightarrow \frac{W}{c\tau +d}.
\end{equation}
since $G$ must be invariant.
That is, the superpotential must be a holomorphic function of the modular weight $-1$.

Note that the K\"ahler potential in $G$ is invariant under the transformation, $\tau \rightarrow -\tau^*$.
Thus, $G$ and the scalar potential $V$ are $CP$-invariant if $|W|^2$ is invariant:
\begin{equation}\label{eq:CP-W}
  W(\tau) \longrightarrow W(-\tau^*) = e^{i\chi} \overline{W(\tau)} ,
\end{equation}
under the $CP$ with $\tau \rightarrow -\tau^*$ including the $CP$ transformation of chiral matter fields
(See Ref.~\cite{Novichkov:2019sqv}.).

Let us study the $A_4$ modular invariant model.
When the modulus-dependent superpotential $W(\tau)$ is generated by some nonperturbative mechanism, the modulus can be stabilized at a potential minimum
(See for early studies \cite{Ferrara:1990ei,Cvetic:1991qm}.).
The superpotential $W(\tau)$ must have the modular weight $-1$ as mentioned above.
However, the $A_4$ modular invariant theory has no modular form of odd weight.
We need some mechanism to compensate the difference of modular weights.

For example, we assume that the condensation $\langle Q \bar Q \rangle \neq 0$ occurs in the hidden sector by strong dynamics such as supersymmetric QCD \cite{Intriligator:1995au}.
When the tree-level superpotential includes the modular-symmetric mass term, $W \sim m(\tau) Q \bar Q$, 
the condensation could lead to nonperturbative superpotential $W\sim m(\tau) \langle Q \bar Q \rangle $, 
and the modular weight $k_m$ of $m(\tau)$ depends on the modular weights $k_Q$ and $k_{\bar Q}$ of  
$Q$ and $\bar Q$, i.e. $k_m= -k_Q -k_{\bar Q} -1$.
Furthermore, when the flavor number of $Q$ and $\bar Q$ is sufficiently small, strong supersymmetric QCD 
dynamics induces the nonperturbative superpotential $W \sim \Lambda(\tau) /{\rm det}(Q \bar Q)$ \cite{Intriligator:1995au}.
We require the modular invariance of $G$ with such nonperturbative superpotential.
Then, the modular weight of $\Lambda(\tau)$ is determined by the modular weights of  
$Q$ and $\bar Q$.\footnote{For example, $\Lambda(\tau)$ can be obtained by the gauge kinetic function $f(\tau)$ like 
$\Lambda(\tau) \sim e^{-af(\tau)}$, and through explicit string calculations in some examples it was found that 
threshold corrections of $f(\tau)$ include modular functions such as the Dedekind $\eta$ function, i.e. $e^{-af(\tau)} \sim \eta(\tau)^{-p}$.
For more details, see Refs.~\cite{Lust:2003ky,Blumenhagen:2006ci} and references therein.}
At any rate, from the bottom-up viewpoint, we assume that the hidden dynamics induces the nonperturbative superpotential 
$W = \Lambda_d g(\tau)$, where $\Lambda_d$ has nonvanishing modular weight originated from  one of the condensation 
$\langle Q \bar Q \rangle $ and the modular weight of $g(\tau)$  is different from $-1$.
The $T$ symmetry can remain, 
because $\langle Q \bar Q \rangle $ can be invariant under $T$ and $\Gamma_N$.

Now, let us assume that  the following superpotential is induced,
\begin{equation}
	\label{eq:W-tau}
	W = \Lambda_d (Y^{(3,4)}_{\bf 1}(\tau))^{-1},
\end{equation}
where $\Lambda_d$ is the dynamical scale which is related to the condensation, e.g. $\Lambda_d = m\langle Q \bar Q \rangle$.
We also assume that $\Lambda_d$ has the modular weight 3.
Note that $|W|^2$ is invariant under $\tau \rightarrow -\tau^*$.

\begin{figure}[h!]
	\begin{tabular}{ccc}
		\begin{minipage}{0.475\linewidth}
			\includegraphics[bb=0 0 400 280,width=\linewidth]{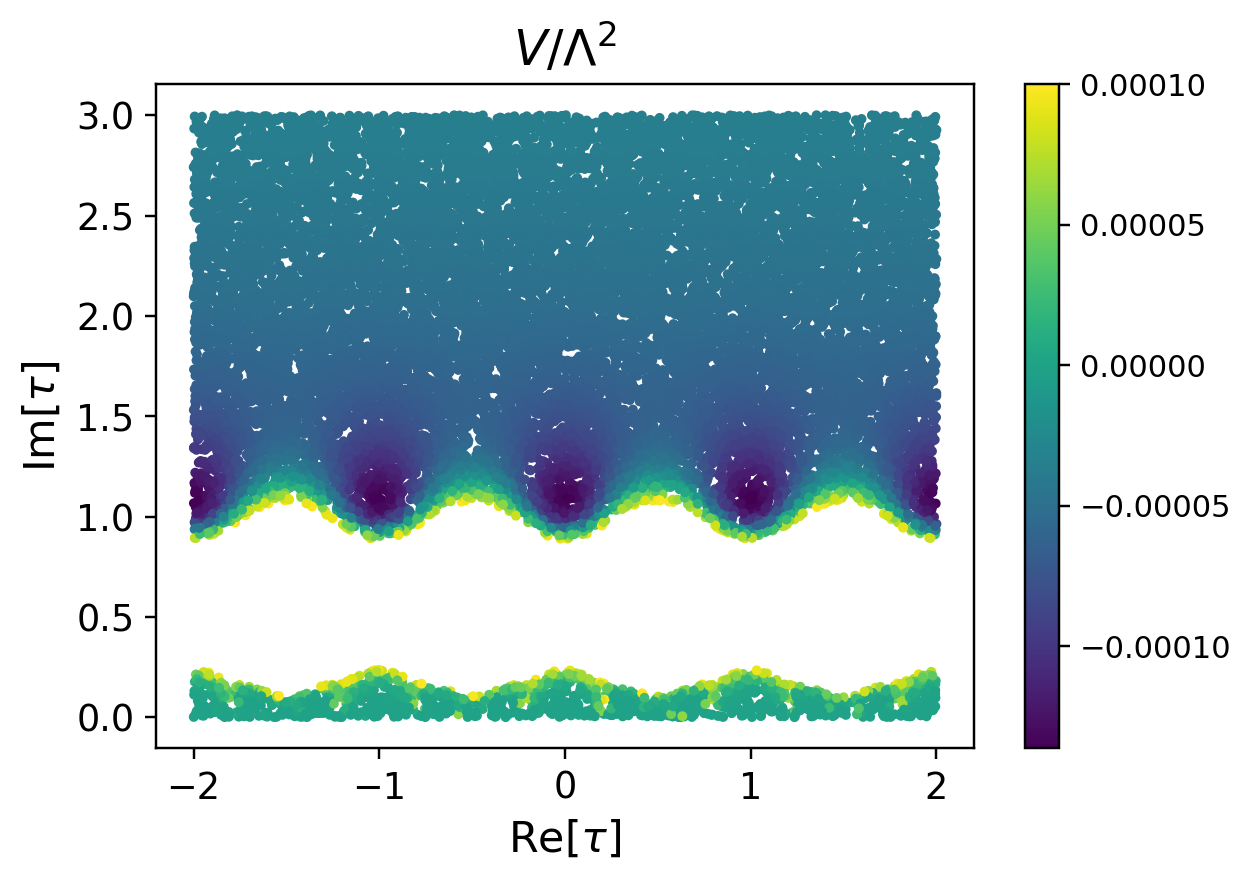}
			 \caption{The contour plot of the scalar potential for $ W$ in Eq.~(\ref{eq:W-tau}).}
			\label{fig:V-min}
		\end{minipage}
		\phantom{=}
		\begin{minipage}{0.475\hsize}
			\includegraphics[bb=0 0 400 280,width=\linewidth]{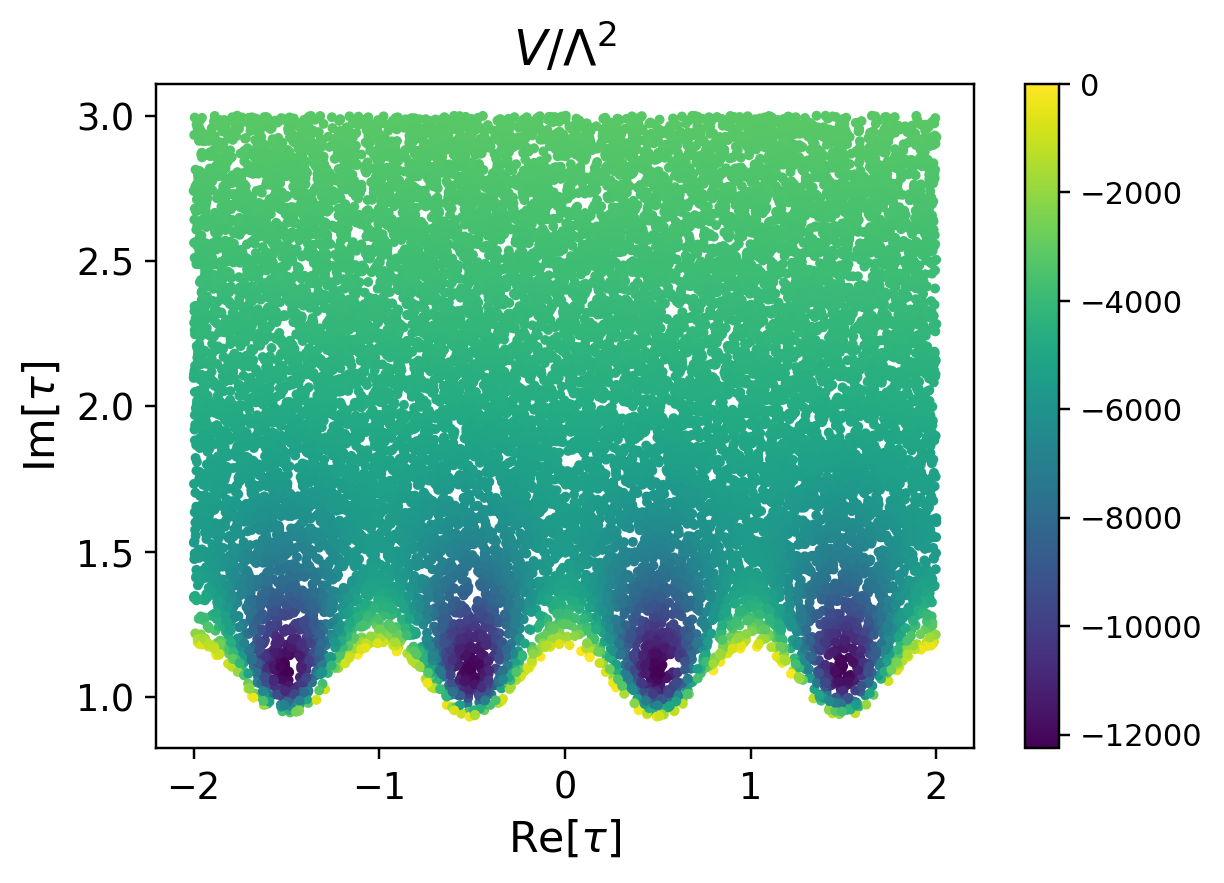}
			\caption{The contour plot of the scalar potential for $ W$ in Eq.~(\ref{eq:W'-tau}).}
			\label{fig:V-min-2}
		\end{minipage}
	\end{tabular}
\end{figure}

We analyze the potential minimum of the scalar potential $V$ with the above ansatz for the superpotential Eq.~(\ref{eq:W-tau}).
The supersymmetric vacuum corresponds to $D_\tau W=0$.
However, there is no solution for $D_\tau W=0$ at a finite value of $\tau$.
There is no supersymmetric vacuum, but there is a nonsupersymmetric vacuum.
Figure \ref{fig:V-min} shows the potential minima at 
\begin{equation}
  \tau_{min} = 1.09i + n,
\end{equation}
where $n$ is integer.
Thus, the stabilized point is ${\rm Re}[\tau]=0$ (mod 1).
$CP$ is not violated at these minima.
This result can be found reasonable by the $q$-expansion of the singlet modular form $Y^{(3,4)}_{\bf 1}(\tau) = 1 + c_1q + c_2q^2+\cdots$, 
where $c_i$ are positive.
The scalar potential $V$ depends on $\cos {2 \pi n{\rm Re}[\tau]}$, 
and the potential becomes minimum at $\cos {2 \pi {\rm Re}[\tau]} = 1$.
Later, we will show illustrating examples in the global supersymmetric limit.

We also assume and analyze the potential minimum of the following superpotential alternatively,
\begin{equation}
	\label{eq:W'-tau}
	W = \Lambda_d (Y^{(3,4)}_{\bf 1}(\tau)),
\end{equation}
where we assume that $\Lambda_d$ has the modular weight $-5$.
Again, there is no supersymmetric vacuum satisfying $D_\tau W =0$ at a finite value of $\tau$.
Figure \ref{fig:V-min-2} shows the potential minima at 
\begin{equation}
  \tau_{min} = 1.09i + p/2,
\end{equation}
where $p$ is odd.
Thus, we have obtained ${\rm Re}[\tau]=1/2$ (mod 1).
This result also can be understood by the $q$-expansion.
The symmetry (\ref{eq:CP-tau}) is violated at this minimum without 
taking account of the $T$ symmetry.
The modular forms have $CP$ phases at ${\rm Re}[\tau]=1/2$  (mod 1).
Such phases may appear as physical $CP$ phases.
However, the symmetry (\ref{eq:CP-tau}) is conserved up to the $T$ symmetry at 
${\rm Re}[\tau]=1/2$ (mod 1).
We will discuss its meaning in the next section.

The above potentials become singular at ${\rm Im}[\tau] \rightarrow 0$ and $\infty$, which are related to
each other by the S transformation.
This is a reasonable behavior from superstring theory, because at the limit ${\rm Im}[\tau] \rightarrow 0$ 
the size of the compact space becomes zero and the gauge couplings diverge.

We can also discuss the $S_3$ invariant model.
The scalar potential $V$ with the following superpotential:
\begin{equation}
	\label{eq:W-tau-S3}
	W = \Lambda_d (Y^{(2,4)}_{\bf 1}(\tau))^{-1},
\end{equation}
has the minimum at ${\rm Re}[\tau]=0$  (mod 1).
On the other hand, the scalar potential $V$ with the following superpotential:
\begin{equation}
	\label{eq:W'-tau-S3}
	W = \Lambda_d (Y^{(2,4)}_{\bf 1}(\tau)),
\end{equation}
has the minimum at ${\rm Re}[\tau]=1/2$  (mod 1).

Similarly, we can discuss the $S_4$ invariant model.
The scalar potential $V$ with the following superpotential:
\begin{equation}
	\label{eq:W-tau-S4}
	W = \Lambda_d (Y^{(4,4)}_{\bf 1}(\tau))^{-1},
\end{equation}
has the minimum at  ${\rm Re}[\tau]=0$  (mod 1), 
while the scalar potential $V$ with the following superpotential:
\begin{equation}
	\label{eq:W'-tau-S4}
	W = \Lambda_d (Y^{(4,4)}_{\bf 1}(\tau)),
\end{equation}
has the minimum ${\rm Re}[\tau]=1/2$  (mod 1).

We have studied the modulus stabilization within the framework of supergravity theory, which is 
proper to study the modulus stabilization because the size of modulus would be large and 
supergravity effects would  not be negligible.
At any rate, we can examine the modulus stabilization within the global supersymmetric theory, 
where the modulus has the K\"ahler  metric 
 \begin{equation}
\frac{|\partial \tau|^2}{(2{\rm Im}[ \tau])^2},
\end{equation}
and the superpotential $W_g$ similar to the above.\footnote{
The superpotential of global supersymmetric theory 
has the vanishing modular weight.}
However, even for such a global supersymmetric theory, 
we obtain the same results that the potential minima correspond to ${\rm Re}[\tau] = 0 $ or $1/2$ (mod 1).
Here we illustrate it by using $q$-expansions.

As mentioned above, the modular form $Y(\tau)$ corresponding to the $\Gamma_N$ trivial singlet can be written by 
\begin{equation}
Y(\tau) = 1 + c_1 q + c_2 q^2 + \cdots,
\end{equation}
where $c_i$ is positive.
For example, we obtain $|q| \sim 10^{-3}$ and $4 \times 10^{-3}$ for ${\rm Im}[\tau]=1$ and $\sqrt{3}/2$.
Thus, we can approximate the above modular form by the first few terms for ${\rm Im}[\tau] \gtrsim 1$.
With such approximation, let us study the following superpotential, 
\begin{equation}
W_g = \Lambda_d (1 + c_1 q + c_2 q^2).
\end{equation}
Then, the scalar potential of the global supersymmetric model is written by 
\begin{equation}
V \sim 16 \pi^2 (\Lambda_d {\rm Im}[\tau])^2 e^{-4\pi {\rm Im}[\tau]}\left( c_1 + 4 c_1c_2 e^{-2\pi {\rm Im}[\tau]}
\cos 2\pi {\rm Re}[\tau]) \right).
\end{equation}
Its minimum is ${\rm Re}[\tau] = 1/2$ (mod 1).

When we assume the following superpotential, 
\begin{equation}
W_g = \frac{\Lambda_d}{1 + c_1 q},
\end{equation}
the corresponding scalar potential is written by 
\begin{equation}\label{eq:Vg-1}
V \sim 16 \pi^2 (\Lambda_d{\rm Im}[\tau])^2 \frac{c_1^2e^{-4\pi {\rm Im}[\tau]}}{ 1 + 4  e^{-2\pi {\rm Im}[\tau]}
\cos ( 2\pi {\rm Re}[\tau]) }.
\end{equation}
Its minimum is ${\rm Re}[\tau] = 0$ (mod 1).
Even if we assume the superpotential, which is a combination of the above, 
\begin{equation}
W_g = c_0 + c_1 q + c_2 q^2 + \frac{b}{1 + c_1' q},
\end{equation}
its scalar potential is written by 
\begin{equation}\label{eq:Vg-2}
V \sim 16 \pi^2 ({\rm Im}[\tau])^2 e^{-4\pi {\rm Im}[\tau]}\frac{B^2+A'\cos (2\pi {\rm Re}[\tau])}{ 1 + A
\cos (2\pi {\rm Re}[\tau]) },
\end{equation}
where $A = 4c_1' e^{-2\pi {\rm Im}[\tau]}$,  $B=c_1 - c_1'b$, 
$A'= 4B(c_2 + c_1c_1')e^{-2\pi {\rm Im}[\tau]}$.
The derivative $\partial V/\partial a$ is proportional to $\sin (2 \pi {\rm Re}[\tau])$.
Thus, the minimum corresponds only to either ${\rm Re}[\tau] = 0$ or $1/2$ depending on parameters.

Furthermore, let us examine the following potential,
\begin{equation}
V=c\cos (2\pi {\rm Re}[\tau]) + \frac{b}{1+c'\cos (2\pi {\rm Re}[\tau])},
\end{equation}
which is essentially a linear combination of the scalar potentials in Eqs.~(\ref{eq:Vg-1}) and (\ref{eq:Vg-2}), 
although it may  not be straightforward to  derive such a potential as the scalar potential of four-dimensional 
supersymmetric theory.
Since $\cos (2\pi {\rm Re}[\tau])$ is originated from $q$, it should have the suppression factor 
such as $e^{-2\pi {\rm Im}[\tau]}\cos (2\pi {\rm Re}[\tau])$.
Thus, we expect $0 < c,c' \ll 1$, while $b$ is a free parameter.
Then, the derivative is obtained by 
\begin{equation}\label{eq:V-derivative}
\frac{\partial V}{\partial a} = 2 \pi \frac{\sin (2\pi {\rm Re}[\tau])}{(1+ \cos (2\pi {\rm Re}[\tau]))^2}
(-c (1+ c'\cos (2\pi {\rm Re}[\tau]))^2 +c'b).
\end{equation}
Obviously, the stationary condition, $(\partial V)/(\partial a)=0$ is satisfied 
at $\sin (2\pi {\rm Re}[\tau]) = 0$, which is the same as the above models.
We may have another solution, 
\begin{equation}
(1+ c'\cos (2\pi {\rm Re}[\tau]))^2 = \frac{c'b}{c}.
\end{equation}
However, note that $c'$ is very small like ${\cal O}(10^{-3})$ or more.
In order to have such a solution, we need fine-tuning, $(c'b)/c = 1 + \varepsilon$.
Then, the solution corresponds to $2c'\cos (2\pi {\rm Re}[\tau])=  \varepsilon$.
It is not clear whether one can realize  such a fine-tuning in a reasonable way.

As results, the $CP$-invariant potential has the minimum at ${\rm Re}[\tau] = 0 $ or $1/2$ (mod 1)
in the simple Ansatz of  potential forms.
We emphasize that the important point is the $T$ transformation in this result: ${\rm Re}[\tau] = {\rm Re}[\tau] + 1$.
It is also important that $\cos (2\pi m{\rm Re}[\tau])$ always appears with the factor $e^{-2\pi m{\rm Im}[\tau]}$ 
in the modular forms.
There is no contribution from large $m$, and the potential is dominated by 
$\cos (2\pi {\rm Re}[\tau])$ .
Thus, the potential must be a function of $\cos (2\pi {\rm Re}[\tau])$.
Then, its derivative is always proportional to $\sin (2\pi {\rm Re}[\tau])$, and 
$(\partial V)/(\partial a)=0$ is satisfied 
at $\sin (2\pi {\rm Re}[\tau]) = 0$.
When the first derivative of generic function may also be proportional to 
a function of  $\cos (2\pi {\rm Re}[\tau])$ like Eq.(\ref{eq:V-derivative}), 
we may need some complication and fine-tuning to realize minima except $\sin (2\pi {\rm Re}[\tau]) = 0$

\section{$CP$ violation in $A_4$ models}

In the previous section, we have shown that the $CP$-invariant scalar potential has the minimum ${\rm Re}[\tau]=0$ or $1/2$ 
(mod 1) by using the simple Ansatz for potential forms.
That is the quite clear result.
It is obvious that the $CP$ is not violated when ${\rm Re}[\tau] = 0 $.
The line  ${\rm Re}[\tau] = 1/2 $ (mod 1) corresponds to the $CP$ symmetric line up to the modular transformation, 
in particular, the $T$ transformation.
The modular forms have nonvanishing phase at ${\rm Re}[\tau]=1/2$ in general
because the modular forms of the level $N$ are expanded by $q^{1/N}$.
We study the meaning of this phase for ${\rm Re}[\tau] = 1/2 $ by using an $A_4$ flavor model in Ref
.~\cite{Kobayashi:2018scp} and discuss a generic aspect.

First, let us study the $A_4$ modular invariant model in Ref.~\cite{Kobayashi:2018scp}.
Table \ref{tb:fields} shows the modular weights and $A_4$ representations of the left-handed leptons and right-handed charged leptons in the model of Ref.~\cite{Kobayashi:2018scp}.
The model in Ref.~\cite{Kobayashi:2018scp} is the global supersymmetric model where the superpotential has the vanishing modular weight.
We rearrange the modular weights of chiral superfields such that the supergravity superpotential has the modular weight $-1$.

\begin{table}[h]
	\centering
	\begin{tabular}{|c||c|c|c|c|} \hline 
		&$L$&$e_R,\mu_R,\tau_R$&$H_u$&$H_d$\\ \hline \hline 
		\rule[14pt]{0pt}{0pt}
		$SU(2)$&$2$&$1$&$2$&$2$\\
		$A_4$&$3$& $1$,\ $1''$,\ $1'$&$1$&$1$\\
		$-k_I$&$-1$&$-1 $&-1/2&-1 \\ \hline
	\end{tabular}
	\caption{
		The charge assignment of $SU(2)$, $A_4$, and the modular weights $-k_I$ for fields.}
	\label{tb:fields}
\end{table}

The Yukawa couplings terms in the superpotential are written by 
\begin{align}
  W_e&=\alpha e_RH_d(LY^{3,2}_{\bf 3})+\beta \mu_RH_d(LY^{3,2}_{\bf 3})+\gamma \tau_RH_d(LY^{3,2}_{\bf 3})~,\label{charged} 
\end{align}
and the Weinberg operator terms are written by 
\begin{align}
  W_\nu&=-\frac{1}{\Lambda}(H_u H_u LLY)_{\bf 1}~,
  \label{Weinberg}
\end{align}
where $\alpha, \beta, \gamma$, and $\Lambda$ are real.
These superpotential terms are $CP$-invariant satisfying Eq.(\ref{eq:CP-W}).
The phase appears only in $q=e^{2\pi i \tau}$ of the modular forms.
The charged lepton mass matrix is written by 
\begin{align}
  \begin{aligned}
  M_E&=v_d{\rm diag}[\alpha, \beta, \gamma]
  \begin{pmatrix}
  Y_1^{3,2} & Y_3^{3,2} & Y_2^{3,2} \\
  Y_2^{3,2} & Y_1^{3,2} & Y_3^{3,2} \\
  Y_3^{3,2} & Y_2^{3,2} & Y_1^{3,2}
  \end{pmatrix}_{RL},
  \end{aligned}\label{eq:CL}
\end{align}
and the neutrino mass matrix is written by 
\begin{align}\label{eq:mass-nu}
  M_\nu=-\frac{v_u^2}{\Lambda}\begin{pmatrix}
  2Y_1^{3,2} & -Y_3^{3,2} & -Y_2^{3,2} \\
  -Y_3^{3,2} & 2Y_2^{3,2} & -Y_1^{3,2} \\
  -Y_2^{3,2} & -Y_1^{3,2} & 2Y_3^{3,2} \end{pmatrix}_{LL},
\end{align}
where we renormalize the coefficients $\alpha, \beta, \gamma$, and $\Lambda$ by normalization factors of kinetic terms,
because such normalization factors are irrelevant to the $CP$ phase.

We set ${\rm Re}[\tau]=1/2$ and the mass matrices have the following phase behavior as shown in Eq.(\ref{q-expansion}):
\begin{align}
  \begin{aligned}
  M_E&= 
  \begin{pmatrix}
  m_{e11} & m_{e12} e^{2i\phi} & m_{e13}e ^{i \phi} \\
  m_{e21}e^{i\phi} & m_{e22} & m_{e23}e^{2 i \phi} \\
  m_{e31} e^{2i\phi } &m_{e32}e^{i\phi} & m_{e33}
  \end{pmatrix} ,
  \end{aligned}\label{eq:ME}
\end{align}
and 
\begin{align}
  \begin{aligned}
  M_\nu&= 
  \begin{pmatrix}
  m_{\nu11} & m_{\nu 12} e^{2i\phi} & m_{\nu13}e ^{i \phi} \\
  m_{\nu 12}e^{2i\phi} & m_{\nu 22}e ^{i \phi} & m_{\nu 23} \\
  m_{\nu 13} e^{i\phi } &m_{\nu 23} & m_{\nu 33}e ^{2i \phi}
  \end{pmatrix} ,
  \end{aligned}\label{eq:Mnu}
\end{align}
where $\phi = \pi/3$, and $m_{eij}$ and $m_{\nu ij}$ are real.
This phase structure is very specific and we can rephase $L_i$ and $E_i$ such that all the phases vanish in these mass matrices.
Thus, there is no physical $CP$ phase.

The above behavior of the $CP$ phase appears not only in the above model.
That is a rather generic property.
Let us study the flavor model with the $\Gamma_N$ flavor modular symmetry.
We use the basis that $\rho(T)$ is represented by a diagonal matrix.
It means that the chiral field $\Phi_i$ transforms 
\begin{equation}
  \Phi_i \rightarrow e^{2\pi i k_i/N} \Phi_i,
\end{equation}
under the $T$ transformation $\tau \rightarrow \tau + 1$, where $k_i$ is an integer because $\rho(T)^N = \mathbb{I}$.
This is the $Z_N$ symmetry.
We study the mass terms in the superpotential,
\begin{equation}\label{eq:mass-W}
  W=M_{ij}\Phi_i \Phi_j.
\end{equation}
These mass terms may originate from the Yukawa coupling terms or Weinberg operator terms in the superpotential.
The mass matrix $M_{ij}$ depends on the modulus $\tau$, and the mass matrix $M_{ij}$ must transform under the $T$ transformation.
At any rate, $M_{ij} \Phi_i \Phi_j$ should be invariant under the $T$ transformation because it is a trivial singlet of $\Gamma_N$.
It transforms as 
 $M_{ij} \rightarrow e^{-2\pi i(k_i+k_j)/N} M_{ij}$ to cancel 
the transformation of $\Phi_i \Phi_j$.
Hence, the mass matrix must have the following form:
\begin{equation}
  M_{ij} = m_{ij}(q) q^{- (k_i+k_j)/N} = m_{ij}(q) e^{- 2\pi i (k_i+k_j)\tau/N},
\end{equation}
where $m_{ij}(q)$ include only integer powers of $q$, i.e. $q^n$.
When ${\rm Re}[\tau] = 1/2$, the phase behavior of the mass matrix must be written by 
\begin{equation}
  M_{ij} = \tilde m_{ij}e^{-\pi i(k_i + k_j)/N},
\end{equation}
where $\tilde m_{ij} = m_{ij} e^{-2\pi (k_i + k_j){\rm Im}[\tau]/N}$ and they are real.
Such phases can be canceled by rephasing $\Phi_i \rightarrow e^{\pi i k_i/N} \Phi_i$.
Thus, there is no physical $CP$ phase for ${\rm Re}[\tau] =1/2$ as expected.\footnote{A similar result is obtained 
for ${\rm Re}[\tau] =-1/2$.}
If ${\rm Re}[\tau] \neq 1/2$ or $0$ (mod 1), $m_{ij}(q)$ can lead to a physical $CP$ phase, which cannot be 
removed by rephasing fields, although the phases $e^{- 2\pi i (k_i+k_j)\tau/N}$ can be removed.

As a result, the $CP$ conserves in the modular invariant flavor model at ${\rm Re}[\tau] =1/2$ as well as ${\rm Re}[\tau] =0$ (mod 1);
and these values of ${\rm Re}[\tau] $ are realized as the minimum of the modular invariant potential.
The spontaneous $CP$ violation does not occur in general.
An important point of this result is the $T$ invariance.
The $T$ invariance prevents the $CP$ violation.
The spontaneous $CP$ violation may occur if the $T$ symmetry is violated.

For example, some representations of $\Gamma_N$ have ${\rm det}\rho (T) \neq 1$.
Fermions with such representations can lead to anomalies \cite{Araki:2008ek}.\footnote{
  See also \cite{Kariyazono:2019ehj}.}
Nonperturbative effects can induce breaking terms for anomalous symmetries.
If the $T$ symmetry is anomalous, mass terms corresponding to nontrivial singlets can appear in the superpotential, e.g.
\begin{equation}
  W=(M_{ij}^{(0)}+ M_{ij}^{(1)}+ M_{ij}^{(2)} \cdots )\Phi_i \Phi_j,
  \end{equation}
  where $M_{ij}^{(0)} = M_{ij}$ in the mass term (\ref{eq:mass-W}). 
(Note that this superpotential conserves the $CP$ symmetry in the sense (\ref{eq:CP-W}), while it breaks 
the $T$ symmetry.)
  The mass matrix $M_{ij}^{(k)}$ has the $Z_N$ charge different from $M^{(0)}$ by $k$, and 
  transforms as 
  \begin{equation}
  \frac{M_{ij}^{(k)}}{M_{ij}^{(0)}} \rightarrow e^{2 \pi i k/N}\frac{M_{ij}^{(k)}}{M_{ij}^{(0)}}.
\end{equation}
We cannot cancel the phases in the mass matrices by rephasing and the physical $CP$ phase appears in this case 
for a generic value of ${\rm Re}[\tau] $ including $1/2$ except $0$.
Thus, the violation of the $T$ symmetry is important to realize the nonvanishing physical $CP$ phase.

The $T$ symmetry in the above $A_4$ model is anomaly free.
The $T$ symmetry becomes anomalous if we consider a specific assignment: for example, ${\bf 1}$ and two ${\bf 1}'$ to the three families of right-handed leptons.
However, it is another problem whether such an assignment can lead to realistic masses and mixing angles.

\section{Conclusion}

We have studied the $CP$ violation through the modulus stabilization by using the simple Anstaz of potential forms.
The $CP$-invariant potential has the minimum at ${\rm Re}[\tau] = 0 $ or $1/2$ (mod 1).
This result is very strong.
It is clear that the $CP$ is not violated at ${\rm Re}[\tau] = 0 $.
However, some modular forms have nonvanishing phases at ${\rm Re}[\tau] = 1/2 $ (mod 1).
We have studied explicitly the $A_4$ flavor model at ${\rm Re}[\tau] = 1/2 $.
This $A_4$ model has a specific structure in the $CP$ phase.
The phases of the modular forms at ${\rm Re}[\tau] = 1/2 $ do not appear as the physical $CP$ phase.
This behavior is not a special case unique to this model.
It is a rather generic property of the $CP$-invariant and modular invariant flavor models.
In particular, the $T$ transformation is important.
The scalar potential has the periodicity $\tau \sim \tau + 1$ since the potential is a trivial singlet of $T$.
Such periodicity leads to the strong result: the minimum is realized at ${\rm Re}[\tau] = 0 $ or $1/2$ (mod 1).
The modular forms have phases at ${\rm Re}[\tau] = 1/2$.
The $T$ invariance, the $Z_N$ symmetry, leads to a phase behavior such that
phases can be canceled by rephasing of the fields and the physical $CP$ phase does not appear.

One needs violation of $T$ symmetry to realize the  $CP$ violation.
For example, anomaly of the $T$ symmetry may lead to the spontaneous $CP$ violation by nonperturbative effects.

\vspace{1.5 cm}
\noindent
{\large\bf Acknowledgement}\\

This work is supported by MEXT KAKENHI Grant Number JP19H04605 (TK), and 
JSPS Grants-in-Aid for Scientific Research 18J11233 (THT).
The work of YS is supported by JSPS KAKENHI Grant Number JP17K05418 and Fujyukai Foundation.

\appendix
\section*{Appendix}

\section{Modular forms for $A_4$}

The modular forms of the level 3 and weight 4 are obtained by products of the modular forms of the weight 2 \cite{Feruglio:2017spp}.
There are five modular forms of the weight 4, and they correspond to 
${\bf 1}$, ${\bf 1}'$, and ${\bf 3}$ of $A_4$,
\begin{equation}
Y_{\bf 1}^{3,4} = (Y_1^{3,2})^2 + 2Y_2^{3,2}Y_3^{3,2}, \qquad 
Y_{{\bf 1}'}^{3,4} = (Y_3^{3,2} )^2 + 2Y_1^{3,2}Y_2^{3,2},
\end{equation}
\begin{align}
Y^{3,4}_{\bf 3}=\begin{pmatrix}Y_{{\bf 3 }1}^{3,4}(\tau)\\
Y_{{\bf 3 }2}^{3,4}(\tau)\\
Y_{{\bf 3 }3}^{3,4}(\tau)\end{pmatrix}=
\begin{pmatrix}
(Y_1^{3,2})^2 - Y_2^{3,2}Y_3^{3,2} \\
(Y_3^{3,2})^2 - Y_1^{3,2}Y_2^{3,2} \\
(Y_2^{3,2})^2 - Y_1^{3,2}Y_3^{3,2}\end{pmatrix}.
\end{align}

\section{Modular forms for $S_4$}

The modular forms of the level 4 and the weight 2 are written by \cite{Penedo:2018nmg},
\begin{align}
	\begin{aligned}
	Y_1(\tau) &= Y(1,1,\omega,\omega^2,\omega,\omega^2|\tau), \\ 
	Y_2(\tau) &= Y(1,1,\omega^2,\omega,\omega^2,\omega|\tau), \\
	Y_3(\tau) &= Y(1,-1,-1,-1,1,1|\tau), \\
	Y_4(\tau) &= Y(1,-1,-\omega^2,-\omega,\omega^2,\omega|\tau), \\
	Y_5(\tau) &= Y(1,-1,-\omega,-\omega^2,\omega,\omega^2|\tau), 
	\end{aligned}\label{eq:Y12345}
\end{align}
where 
\begin{eqnarray}
	Y(a_1,a_2,a_3,a_4,a_5,a_6 \ \tau) &=& 
	a_1 \frac{\eta'(\tau +1/2)}{\eta(\tau +1/2)} +4a_2 \frac{\eta'(4\tau )}{\eta(4\tau )} \nonumber \\
	& &+\frac14 \sum_{m=0}^3a_{m+3} \frac{\eta'((\tau +m) /4)}{\eta((\tau +m)/4 )}.
\end{eqnarray}
These five modular forms correspond to ${\bf 2}$ and ${\bf 3}'$ representations under $\Gamma_4 \simeq S_4$
\begin{equation}
	Y_{S_4 {\bf 2}}(\tau) =\left(
	\begin{array}{c}
	Y_1(\tau) \\
	Y_2(\tau)
	\end{array}
	\right), \qquad Y_{S_4 {\bf 3}'}(\tau) =\left(
	\begin{array}{c}
	Y_3(\tau) \\
	Y_4(\tau) \\
	Y_5(\tau)
	\end{array}
	\right).
\end{equation}
They represent the generators, $S$ and $T$ as 
\begin{equation}
	\rho(S)=\left(
	\begin{array}{cc}
	0 & \omega \\
	\omega^2 & 0
	\end{array}\right), \qquad
	\rho(T)=\left(
	\begin{array}{cc}
	0 & 1 \\
	1 & 0
	\end{array}\right),
\end{equation}
for ${\bf 2}$, and 
\begin{equation}
	\rho(S)=-\frac13\left(
	\begin{array}{ccc}
	-1 & 2\omega^2 & 2 \omega \\
	2\omega & 2 & -\omega^2 \\
	2\omega^2 & -\omega & 2
	\end{array}\right), \qquad
	\rho(T)=
	-\frac13\left(
	\begin{array}{ccc}
	-1 & 2\omega & 2 \omega^2 \\
	2\omega & 2\omega^2 & -1 \\
	2\omega^2 & -1 & 2\omega
	\end{array}\right),
\end{equation}
for ${\bf 3}'$.
These are not symmetric.
The modular form of weight 4 corresponding to the $S_4$ trivial singlet is written by 
\begin{equation}
Y^{4,4}_{\bf 1} = Y_1 Y_2.
\end{equation}
Note that the trivial singlet is obviously symmetric.



\begin{thebibliography}{99}







\bibitem{Altarelli:2010gt}
G.~Altarelli and F.~Feruglio,
Rev.\ Mod.\ Phys.\ {\bf 82} (2010) 2701
[arXiv:1002.0211 [hep-ph]].



\bibitem{Ishimori:2010au}
H.~Ishimori, T.~Kobayashi, H.~Ohki, Y.~Shimizu, H.~Okada and M.~Tanimoto,
Prog.\ Theor.\ Phys.\ Suppl.\ {\bf 183} (2010) 1
[arXiv:1003.3552 [hep-th]].



\bibitem{Ishimori:2012zz}
H.~Ishimori, T.~Kobayashi, H.~Ohki, H.~Okada, Y.~Shimizu and M.~Tanimoto,
Lect.\ Notes Phys.\ {\bf 858} (2012) 1, Springer.

\bibitem{Hernandez:2012ra}
D.~Hernandez and A.~Y.~Smirnov,
Phys.\ Rev.\ D {\bf 86} (2012) 053014
[arXiv:1204.0445 [hep-ph]].

\bibitem{King:2013eh}
S.~F.~King and C.~Luhn,
Rept.\ Prog.\ Phys.\ {\bf 76} (2013) 056201
[arXiv:1301.1340 [hep-ph]].

\bibitem{King:2014nza} 
S.~F.~King, A.~Merle, S.~Morisi, Y.~Shimizu and M.~Tanimoto,
New J.\ Phys.\ {\bf 16}, 045018 (2014)
[arXiv:1402.4271 [hep-ph]].


\bibitem{Tanimoto:2015nfa}
M.~Tanimoto,
AIP Conf.\ Proc.\ {\bf 1666} (2015) 120002.

\bibitem{King:2017guk}
S.~F.~King,
Prog.\ Part.\ Nucl.\ Phys.\ {\bf 94} (2017) 217
[arXiv:1701.04413 [hep-ph]].

\bibitem{Petcov:2017ggy}
S.~T.~Petcov,
Eur.\ Phys.\ J.\ C {\bf 78} (2018) no.9, 709
[arXiv:1711.10806 [hep-ph]].

 


\bibitem{Kobayashi:2006wq}
T.~Kobayashi, H.~P.~Nilles, F.~Ploger, S.~Raby and M.~Ratz,
Nucl.\ Phys.\ B {\bf 768}, 135 (2007)
[hep-ph/0611020].


\bibitem{Kobayashi:2004ya}
T.~Kobayashi, S.~Raby and R.~J.~Zhang,
Nucl.\ Phys.\ B {\bf 704}, 3 (2005)
[hep-ph/0409098].


\bibitem{Ko:2007dz}
P.~Ko, T.~Kobayashi, J.~h.~Park and S.~Raby,
Phys.\ Rev.\ D {\bf 76}, 035005 (2007)
Erratum: [Phys.\ Rev.\ D {\bf 76}, 059901 (2007)]
[arXiv:0704.2807 [hep-ph]].

\bibitem{Beye:2014nxa} 
F.~Beye, T.~Kobayashi and S.~Kuwakino,
Phys.\ Lett.\ B {\bf 736}, 433 (2014)
[arXiv:1406.4660 [hep-th]].


\bibitem{Abe:2009vi}
H.~Abe, K.~S.~Choi, T.~Kobayashi and H.~Ohki,
Nucl.\ Phys.\ B {\bf 820}, 317 (2009)
[arXiv:0904.2631 [hep-ph]].






\bibitem{Lauer:1989ax} 
 J.~Lauer, J.~Mas and H.~P.~Nilles,
 Phys.\ Lett.\ B {\bf 226}, 251 (1989).
%
 Nucl.\ Phys.\ B {\bf 351}, 353 (1991).
 
\bibitem{Lerche:1989cs} 
 W.~Lerche, D.~Lust and N.~P.~Warner,
 Phys.\ Lett.\ B {\bf 231}, 417 (1989).

\bibitem{Ferrara:1989qb} 
 S.~Ferrara, .D.~Lust and S.~Theisen,
 Phys.\ Lett.\ B {\bf 233}, 147 (1989).
 
\bibitem{Kobayashi:2017dyu} 
 T.~Kobayashi and S.~Nagamoto,
 Phys.\ Rev.\ D {\bf 96}, no. 9, 096011 (2017)
 [arXiv:1709.09784 [hep-th]].
 
\bibitem{Kobayashi:2018rad} 
 T.~Kobayashi, S.~Nagamoto, S.~Takada, S.~Tamba and T.~H.~Tatsuishi,
 Phys.\ Rev.\ D {\bf 97}, no. 11, 116002 (2018)
 [arXiv:1804.06644 [hep-th]].
 
\bibitem{Baur:2019kwi} 
 A.~Baur, H.~P.~Nilles, A.~Trautner and P.~K.~S.~Vaudrevange,
 Phys.\ Lett.\ B {\bf 795}, 7 (2019)
 [arXiv:1901.03251 [hep-th]]; 
 %
 arXiv:1908.00805 [hep-th].
 
\bibitem{Kariyazono:2019ehj} 
 Y.~Kariyazono, T.~Kobayashi, S.~Takada, S.~Tamba and H.~Uchida,
 Phys.\ Rev.\ D {\bf 100}, no. 4, 045014 (2019)
 [arXiv:1904.07546 [hep-th]].


\bibitem{deAdelhartToorop:2011re} 
R.~de Adelhart Toorop, F.~Feruglio and C.~Hagedorn,
Nucl.\ Phys.\ B {\bf 858}, 437 (2012)
[arXiv:1112.1340 [hep-ph]].


\bibitem{Feruglio:2017spp}
F.~Feruglio,
arXiv:1706.08749 [hep-ph].

\bibitem{Criado:2018thu}
J.~C.~Criado and F.~Feruglio,
SciPost Phys.\ {\bf 5} (2018) no.5, 042
[arXiv:1807.01125 [hep-ph]].



\bibitem{Kobayashi:2018scp}
T.~Kobayashi, N.~Omoto, Y.~Shimizu, K.~Takagi, M.~Tanimoto and T.~H.~Tatsuishi,
JHEP {\bf 1811} (2018) 196
[arXiv:1808.03012 [hep-ph]].

\bibitem{Kobayashi:2018vbk}
T.~Kobayashi, K.~Tanaka and T.~H.~Tatsuishi,
Phys.\ Rev.\ D {\bf 98} (2018) no.1, 016004
[arXiv:1803.10391 [hep-ph]].


\bibitem{Penedo:2018nmg}
J.~T.~Penedo and S.~T.~Petcov,
Nucl.\ Phys.\ B {\bf 939} (2019) 292
[arXiv:1806.11040 [hep-ph]].

\bibitem{Novichkov:2018nkm}
P.~P.~Novichkov, J.~T.~Penedo, S.~T.~Petcov and A.~V.~Titov,
JHEP {\bf 1904} (2019) 174
[arXiv:1812.02158 [hep-ph]].

\bibitem{Kobayashi:2018bff}
T.~Kobayashi and S.~Tamba,
Phys.\ Rev.\ D {\bf 99} (2019) no.4, 046001
[arXiv:1811.11384 [hep-th]].


\bibitem{Liu:2019khw}
X.~G.~Liu and G.~J.~Ding,
JHEP {\bf 1908} (2019) 134
[arXiv:1907.01488 [hep-ph]].


\bibitem{Novichkov:2018ovf}
P.~P.~Novichkov, J.~T.~Penedo, S.~T.~Petcov and A.~V.~Titov,
JHEP {\bf 1904} (2019) 005
[arXiv:1811.04933 [hep-ph]].



\bibitem{deAnda:2018ecu}
F.~J.~de Anda, S.~F.~King and E.~Perdomo,
arXiv:1812.05620 [hep-ph].

\bibitem{Okada:2018yrn}
H.~Okada and M.~Tanimoto,
Phys.\ Lett.\ B {\bf 791} (2019) 54
[arXiv:1812.09677 [hep-ph]].


\bibitem{Kobayashi:2018wkl} 
T.~Kobayashi, Y.~Shimizu, K.~Takagi, M.~Tanimoto, T.~H.~Tatsuishi and H.~Uchida,
Phys.\ Lett.\ B {\bf 794}, 114 (2019)
[arXiv:1812.11072 [hep-ph]].


\bibitem{Novichkov:2018yse}
P.~P.~Novichkov, S.~T.~Petcov and M.~Tanimoto,
Phys.\ Lett.\ B {\bf 793} (2019) 247
[arXiv:1812.11289 [hep-ph]].



\bibitem{Ding:2019xna}
G.~J.~Ding, S.~F.~King and X.~G.~Liu,
arXiv:1903.12588 [hep-ph].



\bibitem{Nomura:2019jxj} 
 T.~Nomura and H.~Okada,
 Phys.\ Lett.\ B {\bf 797}, 134799 (2019)
 [arXiv:1904.03937 [hep-ph]].


\bibitem{Novichkov:2019sqv} 
 P.~P.~Novichkov, J.~T.~Penedo, S.~T.~Petcov and A.~V.~Titov,
 JHEP {\bf 1907}, 165 (2019)
 [arXiv:1905.11970 [hep-ph]].


\bibitem{Okada:2019uoy} 
 H.~Okada and M.~Tanimoto,
 arXiv:1905.13421 [hep-ph].


\bibitem{deMedeirosVarzielas:2019cyj}
I.~de Medeiros Varzielas, S.~F.~King and Y.~L.~Zhou,
arXiv:1906.02208 [hep-ph].


\bibitem{Nomura:2019yft}
T.~Nomura and H.~Okada,
arXiv:1906.03927 [hep-ph].



\bibitem{Kobayashi:2019rzp} 
 T.~Kobayashi, Y.~Shimizu, K.~Takagi, M.~Tanimoto and T.~H.~Tatsuishi,
 arXiv:1906.10341 [hep-ph].




\bibitem{Okada:2019xqk}
H.~Okada and Y.~Orikasa,
arXiv:1907.04716 [hep-ph].


\bibitem{Kobayashi:2019mna} 
 T.~Kobayashi, Y.~Shimizu, K.~Takagi, M.~Tanimoto and T.~H.~Tatsuishi,
 arXiv:1907.09141 [hep-ph].


\bibitem{Ding:2019zxk} 
 G.~J.~Ding, S.~F.~King and X.~G.~Liu,
 arXiv:1907.11714 [hep-ph].
 
\bibitem{Okada:2019mjf} 
 H.~Okada and Y.~Orikasa,
 arXiv:1907.13520 [hep-ph].
 

\bibitem{King:2019vhv} 
 S.~F.~King and Y.~L.~Zhou,
 arXiv:1908.02770 [hep-ph].
 
 \bibitem{Nomura:2019lnr}
 T.~Nomura, H.~Okada and O.~Popov,
 arXiv:1908.07457 [hep-ph].
 
 \bibitem{Okada:2019lzv}
 H.~Okada and Y.~Orikasa,
 arXiv:1908.08409 [hep-ph].
 
 \bibitem{Criado:2019tzk}
 J.~C.~Criado, F.~Feruglio and S.~J.~D.~King,
 arXiv:1908.11867 [hep-ph].


\bibitem{Gui-JunDing:2019wap} 
 Gui-Jun~Ding, S.~F.~~King, X.~G.~Liu and J.~N.~Lu,
 arXiv:1910.03460 [hep-ph].

\bibitem{Zhang:2019ngf} 
  D.~Zhang,
  arXiv:1910.07869 [hep-ph].



\bibitem{Wang:2019ovr} 
  X.~Wang and S.~Zhou,
  arXiv:1910.09473 [hep-ph].





\bibitem{Ferrara:1989bc} 
 S.~Ferrara, D.~Lust, A.~D.~Shapere and S.~Theisen,
 Phys.\ Lett.\ B {\bf 225}, 363 (1989).
 
\bibitem{Derendinger:1991hq} 
 J.~P.~Derendinger, S.~Ferrara, C.~Kounnas and F.~Zwirner,
 Nucl.\ Phys.\ B {\bf 372}, 145 (1992).
 
\bibitem{Ibanez:1992hc} 
 L.~E.~Ibanez and D.~Lust,
 Nucl.\ Phys.\ B {\bf 382}, 305 (1992)
 [hep-th/9202046].
 
\bibitem{Kobayashi:2016ovu} 
 T.~Kobayashi, S.~Nagamoto and S.~Uemura,
 PTEP {\bf 2017}, no. 2, 023B02 (2017)
 [arXiv:1608.06129 [hep-th]].
 
\bibitem{Ferrara:1990ei} 
 S.~Ferrara, N.~Magnoli, T.~R.~Taylor and G.~Veneziano,
 Phys.\ Lett.\ B {\bf 245}, 409 (1990).
 
\bibitem{Cvetic:1991qm} 
 M.~Cvetic, A.~Font, L.~E.~Ibanez, D.~Lust and F.~Quevedo,
 Nucl.\ Phys.\ B {\bf 361}, 194 (1991).
 

\bibitem{Kobayashi:2016mzg} 
 T.~Kobayashi, D.~Nitta and Y.~Urakawa,
 JCAP {\bf 1608}, no. 08, 014 (2016)
 [arXiv:1604.02995 [hep-th]].



\bibitem{Kobayashi:2019xvz} 
 T.~Kobayashi, Y.~Shimizu, K.~Takagi, M.~Tanimoto and T.~H.~Tatsuishi,
 arXiv:1909.05139 [hep-ph].





\bibitem{Green:1987mn} 
 M.~B.~Green, J.~H.~Schwarz and E.~Witten,
 Cambridge, Uk: Univ. Pr. ( 1987) 596 P. ( Cambridge Monographs On Mathematical Physics)

\bibitem{Strominger:1985it} 
 A.~Strominger and E.~Witten,
 Commun.\ Math.\ Phys.\ {\bf 101}, 341 (1985).


\bibitem{Dine:1992ya} 
 M.~Dine, R.~G.~Leigh and D.~A.~MacIntire,
 Phys.\ Rev.\ Lett.\ {\bf 69}, 2030 (1992)
 [hep-th/9205011].

\bibitem{Choi:1992xp} 
 K.~w.~Choi, D.~B.~Kaplan and A.~E.~Nelson,
 Nucl.\ Phys.\ B {\bf 391}, 515 (1993)
 [hep-ph/9205202].

\bibitem{Lim:1990bp} 
 C.~S.~Lim,
 Phys.\ Lett.\ B {\bf 256}, 233 (1991).

\bibitem{Kobayashi:1994ks} 
 T.~Kobayashi and C.~S.~Lim,
 Phys.\ Lett.\ B {\bf 343}, 122 (1995)
 [hep-th/9410023].





\bibitem{Acharya:1995ag} 
 B.~S.~Acharya, D.~Bailin, A.~Love, W.~A.~Sabra and S.~Thomas,
 Phys.\ Lett.\ B {\bf 357}, 387 (1995)
 Erratum: [Phys.\ Lett.\ B {\bf 407}, 451 (1997)]
 [hep-th/9506143].


\bibitem{Dent:2001cc} 
 T.~Dent,
 Phys.\ Rev.\ D {\bf 64}, 056005 (2001)
 [hep-ph/0105285].


\bibitem{Khalil:2001dr} 
  S.~Khalil, O.~Lebedev and S.~Morris,
  Phys.\ Rev.\ D {\bf 65}, 115014 (2002)
  [hep-th/0110063].


\bibitem{Giedt:2002ns} 
 J.~Giedt,
 Mod.\ Phys.\ Lett.\ A {\bf 17}, 1465 (2002)
 [hep-ph/0204017].



\bibitem{Feruglio:2012cw} 
 F.~Feruglio, C.~Hagedorn and R.~Ziegler,
 JHEP {\bf 1307}, 027 (2013)
 [arXiv:1211.5560 [hep-ph]].

\bibitem{Holthausen:2012dk} 
 M.~Holthausen, M.~Lindner and M.~A.~Schmidt,
 JHEP {\bf 1304}, 122 (2013)
 [arXiv:1211.6953 [hep-ph]].

\bibitem{Chen:2014tpa} 
 M.~C.~Chen, M.~Fallbacher, K.~T.~Mahanthappa, M.~Ratz and A.~Trautner,
 Nucl.\ Phys.\ B {\bf 883}, 267 (2014)
 [arXiv:1402.0507 [hep-ph]].




\bibitem{Intriligator:1995au} 
  K.~A.~Intriligator and N.~Seiberg,
  Nucl.\ Phys.\ Proc.\ Suppl.\  {\bf 45BC}, 1 (1996)
  [Subnucl.\ Ser.\  {\bf 34}, 237 (1997)]
  [hep-th/9509066].



\bibitem{Lust:2003ky} 
  D.~Lust and S.~Stieberger,
  Fortsch.\ Phys.\  {\bf 55}, 427 (2007)
  [hep-th/0302221].


\bibitem{Blumenhagen:2006ci} 
  R.~Blumenhagen, B.~Kors, D.~Lust and S.~Stieberger,
  Phys.\ Rept.\  {\bf 445}, 1 (2007)
  [hep-th/0610327].


\bibitem{Araki:2008ek} 
 T.~Araki, T.~Kobayashi, J.~Kubo, S.~Ramos-Sanchez, M.~Ratz and P.~K.~S.~Vaudrevange,
 Nucl.\ Phys.\ B {\bf 805}, 124 (2008)
 [arXiv:0805.0207 [hep-th]].



\end{thebibliography}
\end{document}